\documentclass[]{aa}
\usepackage{psfig}
\begin{document}

\title{Photometric observations of distant active comets\thanks{Based
on observations taken at the German-Spanish
Astronomical Centre, Calar Alto, operated by the Max-Planck-Instiute for
Astronomy, Heidelberg, jointly with the Spanish National Commission
for Astronomy}}

\author{Gy. Szab\'o\inst{1} \and B. Cs\'ak\inst{1,2} \and K. S\'arneczky\inst{3}
\and L.L. Kiss\inst{1}}

\institute{
Department of Experimental Physics \& Astronomical Observatory,
University of Szeged,
H-6720 Szeged, D\'om t\'er 9., Hungary \and
Department of Optics \& Quantum Electronics \& Astronomical
Observatory, University of Szeged,
POB 406, H-6701 Szeged, Hungary \and
 Department of Physical Geography, ELTE University, H-1088 Budapest,
Ludovika t\'er 2., Hungary}

\titlerunning{CCD photometry of distant active comets}
\authorrunning{Szab\'o et al.}
\offprints{L.L. Kiss, \email{l.kiss@physx.u-szeged.hu}}
\date{}

\abstract{
We present CCD VR$_C$ observations of 6 distant comets located at
heliocentric distances of 3.4--7.2 AU. Time-series data were
obtained on three nights in July, 2000 covering 16 hours.
Each comet was observed after the perihelion,
when a lower activity was expected. Contrary to expectation, 
we found well-defined circular comae
and extended tails visible out to a few (3-5) arcminutes.
We detected a quasi-cyclic light variation of C/1999 J2, while
C/1999 N4 showed some hints of a more complex variation.
C/2000 K1 was constant to $\pm0.04$ mag during the observing run.
The standard V and R$_C$ data were used to estimate
nuclear diameters, while the colour indices implied a slighly reddish
(V$-$R=0\fm68, C/2000 K1), a neutral (V$-$R=0\fm47, C/1999 N4)
and a slightly bluish (V$-$R=0\fm25, C/1999 J2) coma.
Simple fits of the surface brightness distributions are also
presented enabling order of magnitude estimates of nuclear radii.
Beside the time-series observations, further single-shot
observations of three faint comets are briefly described.
}

\maketitle

\section{Introduction}

\begin{table*}
\caption{The journal of observations. (R -- heliocentric distance; $\Delta$ --
geocentric distance; E -- elongation; $\lambda$ -- ecliptic longitude;
$\beta$ -- ecliptic latitude; $\alpha$ - solar phase angle; aspect data are referred to 2000.0;
$\mu_R$ -- mag/arcsec$^2$ in R-band)}
\begin{center}
\begin{tabular} {lrrlllrrrrrr}
\hline
Date & RA & Decl. & R(AU) & $\Delta$(AU) & E & $\lambda (^\circ)$ & $\beta (^\circ)$ & $\alpha (^\circ)$ & see($^{\prime\prime}$) & sky($\mu_R$) & exp(s)\\
\hline
{\bf C/1999 F2} & & & & & & &\\
2000 June 30 & 17 05 & +30 45 & 5.512 & 4.728 & 137.6 & 205.3 & 63.7 & 7.3 & 2.1 & 20.2 & 240\\
{\bf C/1999 J2} & & & & & & &\\
2000 July 06 & 15 20 & +32 29 & 7.135 & 6.836 & 104.2 & 326.2 & 48.1 & 8.0 & 1.5 & 19.8 & 240\\
2000 July 09 & 15 19 & +32 03 & 7.136 & 6.868 & 101.9 & 326.1 & 47.7 & 8.0 & 2.1 & 19.5 & 240\\
{\bf C/1999 N4}& & & & & & &\\
2000 July 04 & 16 23 & +05 35 & 5.514 & 4.771 & 133.9 & 297.2 & 26.7 & 8.0 & 1.6 & 20.7 & 240\\
2000 July 06 & 16 21 & +05 32 & 5.515 & 4.794 & 132.0 & 297.2 & 26.7 & 8.0 & 1.4 & 19.8 & 240\\
{\bf C/1999 T2} & & & & & & &\\
2000 July 02 & 21 29 & +60 25 & 3.351 & 3.318 & 90.5 & 249.6 & 53.3 & 17.7 & 1.9 & 20.2 & 90\\
{\bf C/2000 H1} & & & & & & &\\
2000 June 30 & 15 27 & +48 18 & 3.890 & 3.643 & 96.4 & 347.4 & 66.7 & 15.1 & 2.1 & 20.2 & 240\\
{\bf C/2000 K1}& & & & & & &\\
2000 July 04 & 16 04 & +13 05 & 6.429 & 5.798 & 125.4 & 304.4 & 33.2 & 7.5 & 1.6 & 20.7 & 180\\
\hline
\end{tabular}
\end{center}
\end{table*}

The overwhelming majority of cometary studies
are based on ground-based or space observations of bright comets
around or near to
their perihelion. Direct imaging reveals the inner
structure of the coma, which usually hides the
nucleus itself. There has been a small number of papers
dealing with distant ($R\geq5$ AU) comets
(e.g. O'Ceallaigh et al. 1995, Lowry et al. 1999), and
consequently, there is a serious lack of information
on the behaviour of these objects. On the other
hand, time-series observations may give constraints on the period
of rotation and related effects (Jewitt 1992).

Optical photometry
of sunlight scattered from the nucleus was first
attempted by Fay \& Wisniewsky (1978) on comet 6P/d'Arrest.
The approach is identical to that used in the asteroid studies.
Further CCD photometry using very small aperture radii has been used
to study the rotation of some comets by Jewitt \& Meech (1985), Jewitt (1990),
Meech et al. (1993, 1997), Licandro et al. (2000a, 2000b).
Licandro et al. (2000a) discussed
in detail rotation and shape models based on
the lightcurves. Additional
reduction has to be done with respect to the seeing effects as the
variation in seeing may cause a virtual light variation
of the inner coma with an amplitude of some tenths of a magnitude.

The nucleus is usually well embedded in the coma and
two basic approaches exist to determine the nuclear diameter.
The most efficient one is the use of high spatial resolution
observations of closely passing comets with the HST
(e.g. Lamy \& T\'oth 1995, Lamy et al. 1998, 1999). The other
commonly used method is observing distant nuclei in quiescence,
when the coma is effectively absent (Luu \& Jewitt 1992, O'Ceallaigh et al.
1995, Boehnhardt et al. 1999). Very recently, the nuclear thermal
radiation of some comets was studied in the infrared with help of
ISOCAM observations, resulting in independent diameter
estimates (Jorda et al. 2000).

The new observing strategies by automatic telescopes (e.g. project LINEAR)
and dedicated large instruments (Hainaut \& Meech 1997)
led to regular discoveries of relatively bright, large perihelion-distance
comets,
often more months or a year before the perihelion passage. Thus,
one can study cometary activity in such regions that were
unavailable even a decade ago. The main aim of our work is
to contribute to this topic with new CCD observations
of three comets located between 5.5--7.2 AU --
C/1999 J2 (Skiff), C/1999 N4 (LINEAR)
and C/2000 K1 (LINEAR). Beside the absolute VR$_C$ photometry we took
time-series observations in order to detect rotational effects.
The paper is organised as follows. The observations are
described in Sect.\ 2, while Sect.\ 3 deals with the
analysis and detailed observational results. A brief discussion
is given in Sect.\ 4.

\section{Observations}

\begin{figure*}
\begin{center}
\leavevmode
\psfig{figure=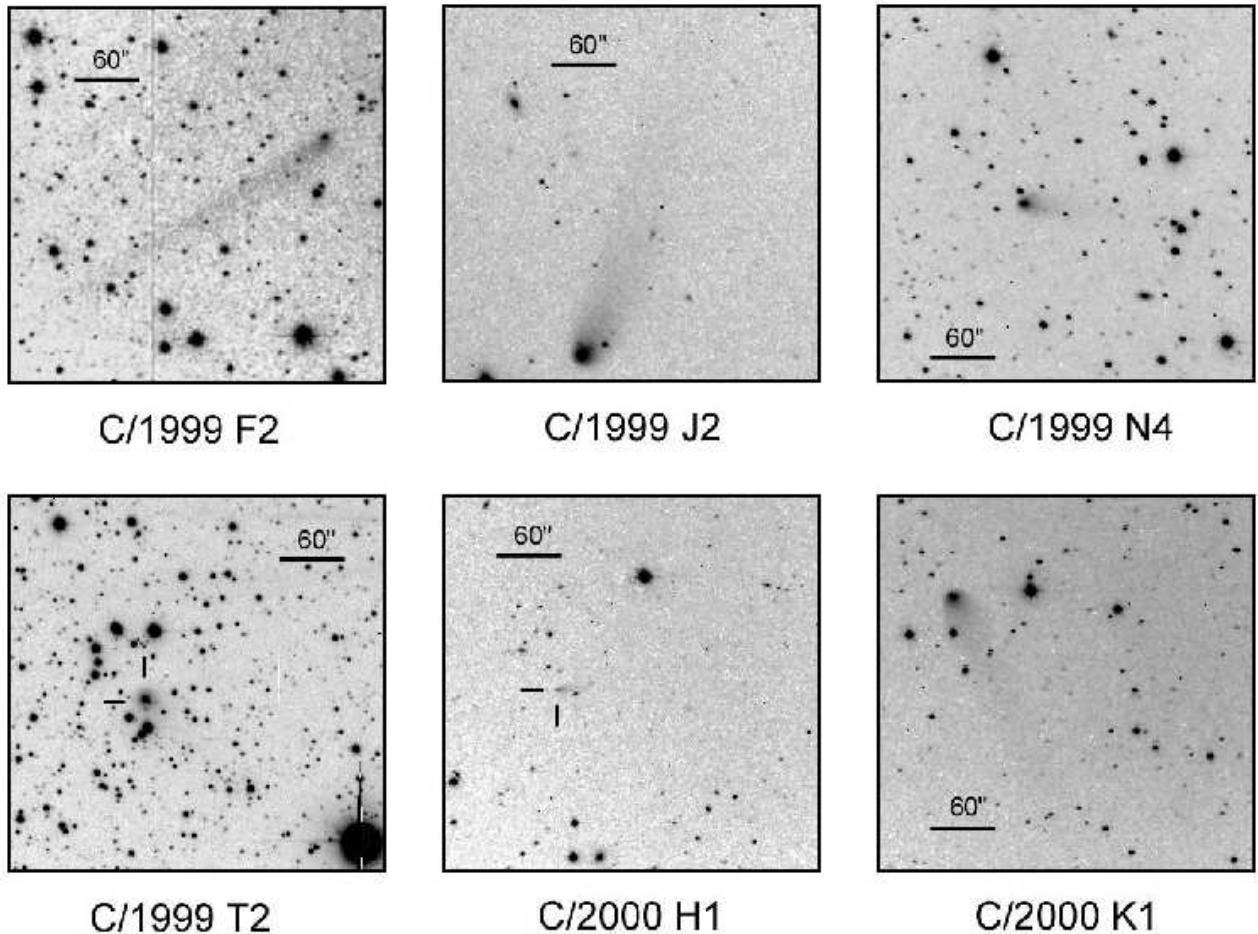,width=\textwidth}
\caption{R-filtered images of the observed comets (C/1999 F2: June 30,
C/1999 J2: July 9,
C/1999 N4: July 4,
C/1999 T2: July 2,
C/2000 H1: June 30,
C/2000 K1: July 4). North is up, west
is left. The images were scaled nonlinearly to enhance the visibility of
diffuse outer regions.}
\end{center}
\label{10298f1}
\end{figure*}

Johnson V and R$_C$ filtered CCD observations were carried
out at Calar Alto Observatory (Spain) on three nights in July, 2000
(4th, 6th and 9th).
The instrument used was the 1.23-m telescope equipped
with the SITe\#2b CCD camera (2048x2048 pixels giving
an angular resolution of 0\farcs49/pixel). The projected
sky area is 16\farcm0$\times$16\farcm0, 10\farcm0$\times$10\farcm0
unvignetted.

The targets were selected following a few
practical restrictions. Since the main aim was to observe
distant comets with small expected coma contributions,
we chose every visible comet at solar distances larger
than 5 AU. The central brightness was restricted to
be brighter than 19.0 mag, since time-series observations
were planned to reveal rotation.
Four comets remained as possible targets, and two further comets
at a solar distance between 3 and 4 AU were added as auxiliary
candidates. Finaly, six comets were examined.

We captured
all of them a few days before starting the comet observations
(on June 30th and July 1st) in order to check which comets would be
well-suited for our purposes. Three of them turned out to be unsuitable
for the detailed analysis, thus we only characterize their comae and tails.
Three objects remained as final targets for the morphological
and photometric studies.

The exposure time was limited by two factors: firstly, the
comets were not allowed to move more than two times the
FWHM of the stellar profiles (varying from night to night) and
secondly, the signal-to-noise (SN) ratio had to be at least 20.
This latter parameter was estimated comparing the peak pixel values
with the sky background during the observations. The adapted
exposure time was 240 seconds. The observing log is summarized in Table\ 1,
listing also the main geometric parameters and sky conditions.

\begin{figure}
\begin{center}
\leavevmode
\psfig{figure=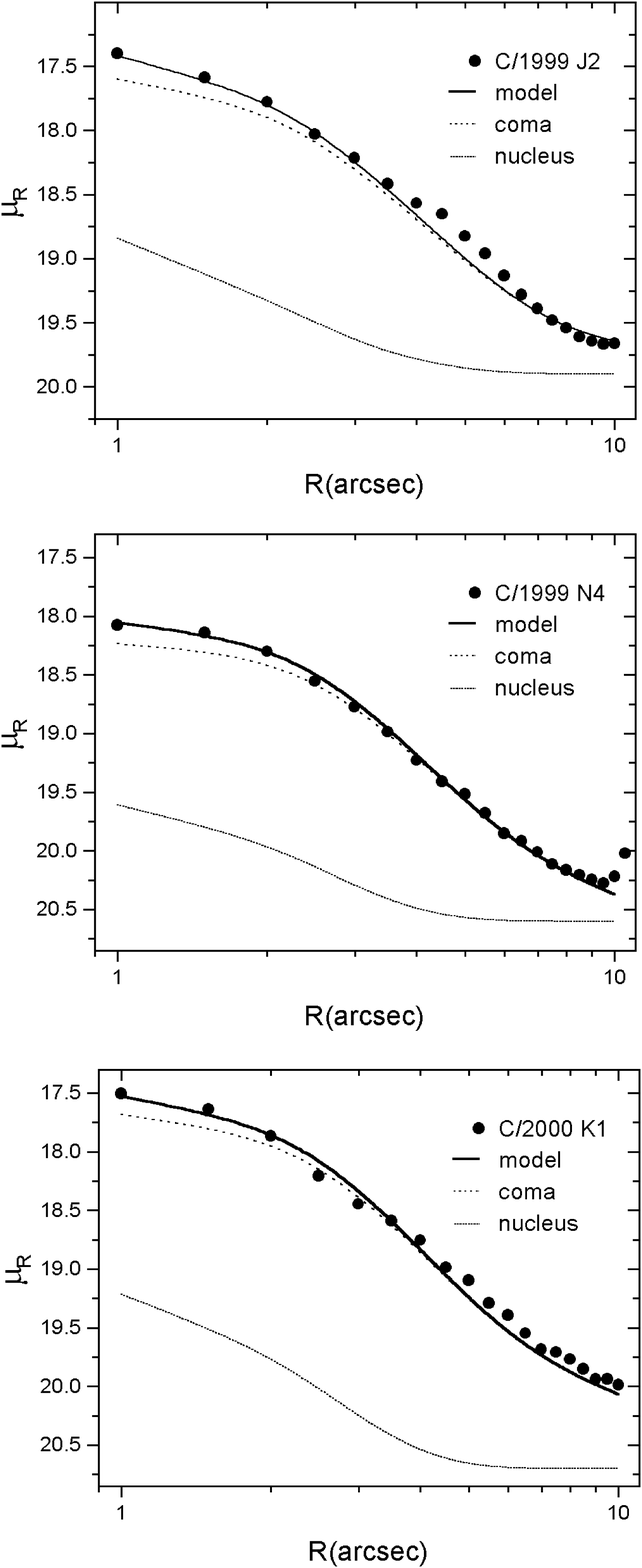,width=8cm}
\caption{R-filtered surface brightness distributions
(in mag/arcsec$^2$) compared with the coma models.
Note the logarithmic scale of the horizontal axes.}
\end{center}
\label{10298f2}
\end{figure}

The CCD frames were reduced with standard tasks
in IRAF. We obtained flat-fields taken during the
evening twillight, and a master flat-field was formed
with the task {\it flatcombine}. The photometric reductions
were done with the corresponding tasks in IRAF/APPHOT.
The trailed images and the presence of faint comae
did not permit the use of psf-photometry. Therefore,
a simple aperture photometry was performed.
The applied differential photometry consisted of
using two stars nearby as comparison and check stars.
The cometary magnitudes are relative to the ensemble mean
of the comparison, thus improving the precision of the
differential data.
We have carefully examined the aperture choice and
a 2\farcs6 diameter was accepted with respect
to the mean FWHM and its doubled value. The time-series
accuracy was estimated by selecting two nearby stars at similar
brightnesses to those of the nuclei for the comparison and
check stars. The usual scatter of comp minus check measurements
was about $\pm$0\fm02.

Since all three nights were photometric, we could make
an absolute ``all-sky'' photometry using photometric
standard stars taken from  Landolt (1992).
The standard field of PG1633+099 used was fairly close to the celestial
positions of the observed comets. This field contains
five standard stars covering V magnitudes between 12\fm969 and 15\fm256
and V$-$R colour between $-$0\fm093 and 0\fm618.
The extinction was monitored by observing an A-type standard at different
air-masses, and the nightly zeropoints of the standard
transformations were determined with the other standards.
The standard
deviation of the linear fits is $\pm$0.02 mag, implying similar
precision for the absolute values.

Another important correction specific for comet photometry
was also applied. Licandro et al. (2000a)
discussed the effects of the varying seeing on the photometry
of blurred diffuse surfaces, such as those observed in comets.
These authors outlined the following method: the actual
seeing is determined in every CCD frame by examining stellar
profiles. Then a seeing-magnitude relation is found
with help of artificially blurred images of a non-variable comet.
In this way the ``seeing-effect'' on the magnitude determination
carried out in the inner coma can be corrected for each observed
frame with varying seeing. This ``seeing-subtraction'' removes
the some part of the atmospheric effect and the remaining variation can be
attributed to the comet itself. The typical corrections
did not exceed 0\fm1. The procedure adds a further
$\sim$0\fm01 noise to the data and including all of the mentioned
uncertainties, the photometric accuracy is estimated to be $\pm$0\fm05.
The data reduction ends with the correction
for the light time\footnote{Original data are available
electronically at CDS, Strasbourg}.

\section{Analysis and results}

In this section we discuss the applied analysis methods and
individual results for the
observed comets. After describing the morphology
we discuss their magnitudes and make an attempt to estimate
the nuclear radii. We also analyze the time series data.

As can be seen in Fig.\ 1,
despite the large solar distances, the observed comets are fairly
active. The general appearance is dominated by a small circular coma
and a faint, considerably long tail.
Although this behaviour is not typical for
Short Period Comets, the Long Period Comets display exactly this
appearance, keeping the level of activity over a very wide range
of distances (Meech 1991). The presented images, that are
single exposures, were enhanced nonlinearly to emphasize
 the faint tail extensions.

\begin{figure}
\begin{center}
\leavevmode
\psfig{figure=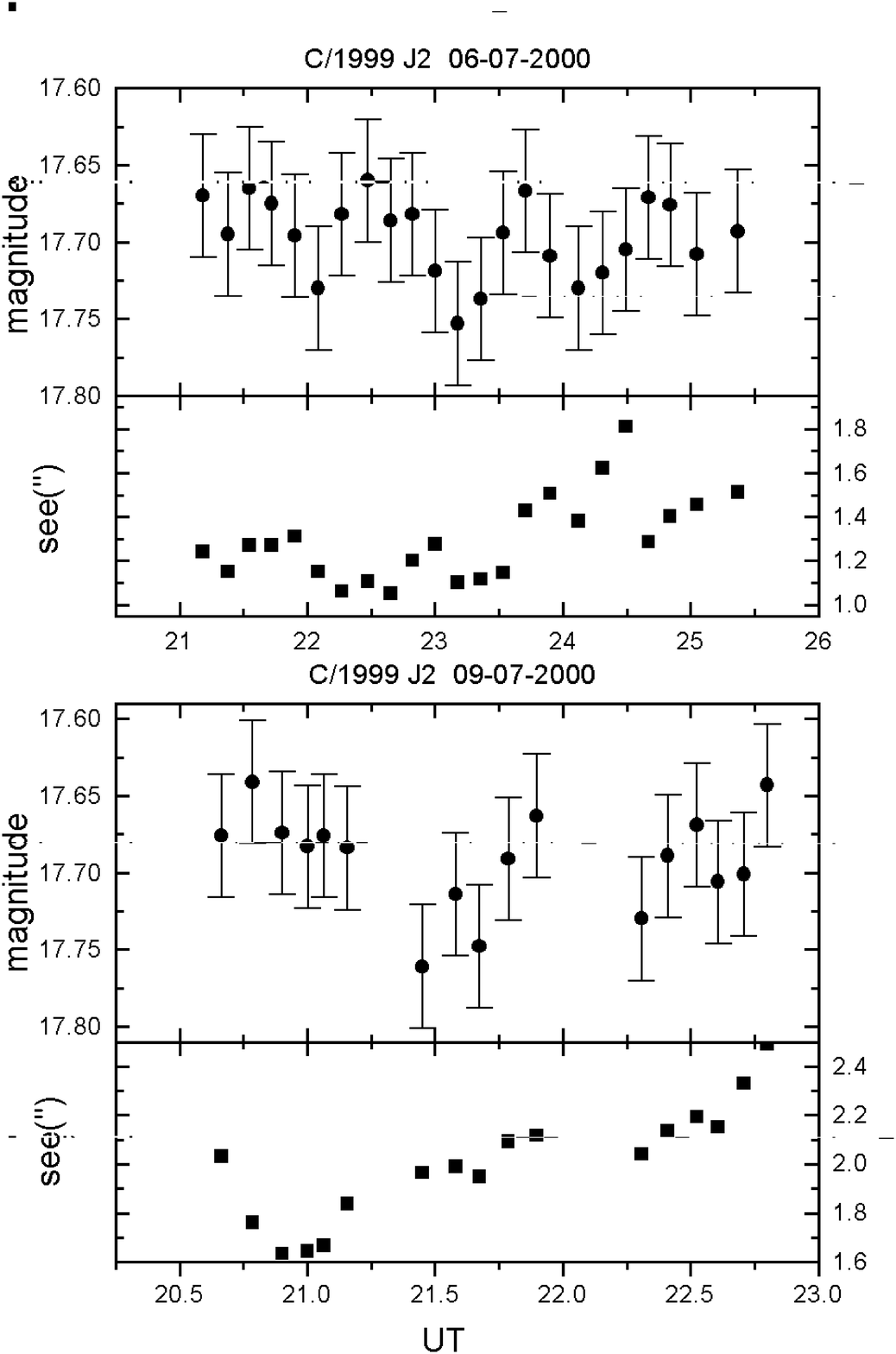,width=8cm}
\caption{The R lightcurves of C/1999 J2 (Skiff) on July 6 and July 9, 2000.
The lower panels show the seeing variations.}
\end{center}
\label{10298f3}
\end{figure}

To derive the physical length of tails one has to estimate their
apparent size. The end of the tail was defined where the
signal-to-noise ratio of the surface brightness was 2 on the
composite images (4 individual exposures co-added).
The apparent tails are visible out to 5\farcm8 (C/1999 F2),
5\farcm2 (C/1999 J2), 3\farcm0 (C/1999 N4), 0\farcm8 (C/1999 T2),
0\farcm3 (C/2000 H1) and 8\farcm9 (C/2000 K1).

The measured position
angles (PA) are the following (antisolar directions in parentheses):
C/1999 F2 -- PA 130$^\circ$ (PA 98$^\circ$);
C/1999 J2 -- PA 18$^\circ$ (PA 80$^\circ$);
C/1999 N4 -- PA 97$^\circ$ (PA 111$^\circ$);
C/1999 T2 -- PA 150$^\circ$ (PA 220$^\circ$);
C/2000 H1 -- PA 90$^\circ$ (PA 76$^\circ$);
C/2000 K1 -- PA 150$^\circ$ (PA 102$^\circ$).
One can find large difference in C/1999 J2 and C/1999 T2
and a smaller one
in C/2000 K1, implying a significant amount of tail curvature in
these comets. This is supported for C/1999 J2 by the antitail
observations by Fukushima et al. (2000) made two months before our
observing run. In the meantime the Earth crossed the orbital
plane and the tail turned significantly. A colour index of C/1999 J2
also supports a dusty tail (see the photometric analysis)
where one may observe apparently long tails if the curvature affects
the projection significantly.
Neglecting this curvature we calculated the true lengths
taking into account the
projection effect. The results are
$19\cdot10^5$ km (C/1999 F2),
$8.5\cdot10^5$ km (C/1999 J2),
$3.2\cdot10^5$ km (C/1999 N4),
$1.2\cdot10^5$ km (C/1999 T2),
$0.5\cdot10^5$ km (C/2000 H1) and
$12\cdot10^5$ km (C/2000 K1).

Based on the test images taken previously on non-photometric nights,
three of the potential targets were
excluded from further observations. The potential targets
C/1999 T2 and C/2000 H1 were not
really ``distant'' object during the observations. Furthermore, C/2000 H1
was quite compact and faint (R=19\fm7, integrated
in the innermost 6$^{\prime\prime}$). C/1999 T2 was bright
(R=15\fm9) with a strongly visible coma and 50$^{\prime\prime}$-long tail.
The integrated R magnitude of C/1999 F2 was 19\fm1 in the inner 6$^{\prime\prime}$,
which is fainter than our practical limit.
We note the impressive cyrrus-like tail as long as 350$^{\prime\prime}$.

The surface brightness profiles (e.g. Jewitt \& Meech 1987, Lowry et al.
1999) were also calculated to examine the coma regularity.
In spherically homogeneous and isotropic cases, the surface brightness
(B) profile bears a simple linear relation with a logarithmic
derivative of $-$1. Interaction between overstreaming matter and the
radiation pressure modifies this value down to about $-$2, according
to the models by Jewitt \& Meech (1987). For our images the profiles
were calculated by a model coma along with the nuclear brightness
estimation.

The model comae were characterized by two free parameters, namely the
slope parameter (logarithmic derivative of the surface brightness)
and coma-to-nucleus brightness ratio. The resultant
structure is a power-function of the radius with negative exponent, while
the nucleus is represented with a delta-function in the center.
To simulate the apparent motion and the effect of seeing, the model
coma was convolved by the motion and the PSF (Lamy et al. 1998).
The latter was determined from individual stellar profiles and its
form was a simple Gaussian. We have performed a least-squares
analysis, where the finally adopted parameters resulted in
an appropriate fit of the observed surface brightness distribution.

The adopted numerical fits of the individual comae are examined
in Fig.\ 2.
A comparison of the measurements, the modelled coma components
and their sums are presented. The measured brightnesses were
averaged in neighbouring rings with a width of $0\farcs5$.
As the background has been taken into account as an additive constant,
the effect has been taken into account in this representation too.
This allows us to show the photometric data with respect to the real limit of
the sky conditions.

Determination of the absolute brightness of the coma allows one to
calculate the apparent magnitude of the core, which results in an
estimate of the diameter of the solid nucleus.
Certainly, as the coma strongly affects the brightness of the nucleus, the
apparent brightness of the nucleus can be determined with quite large errors.
In the case of the present calculations, the coma contamination was about 1.5 -- 2
magnitudes, therefore, the confidence interval
of the estimated diameters may be in the range of half
twice the accepted values.

From the R-band nuclear magnitudes
the mean optical cross sections of the nucleii were determined
using the equation (Eddington 1910)

$p_{\rm R} \overline {C} =
{2.25 \times 10^{22} \pi R^2 \Delta^2 10^{\rm 0.4(m_{\rm Sun}- \overline{m_R})} \over
10^{\rm -0.4 \alpha \beta} }$

\noindent where $p_R$ is the red geometric albedo, $\overline{C}$ is the mean
cross section, $m_{\rm Sun}=-27.96$ mag is the R-band magnitude
of the Sun and $\overline{m_R}$ is the mean R magnitude of the comet, while
$\beta$ is usually assumed to be 0.04 (Luu \& Jewitt 1992).
An important question is the involved
sytematic error, which is very difficult to estimate.
The brightness profile of the coma may be distorted by anisotropic
substructures (e.g. jets, bright patches) blurred by the seeing.
Therefore one can barely calculate the brightness of the nuclei in the case of
ground-based observations, and the resulted nuclear diameters are
often overestimated (I. T\'oth, personal communication).

The remarks on the individual comets are as follows.

\begin{figure}
\begin{center}
\leavevmode
\psfig{figure=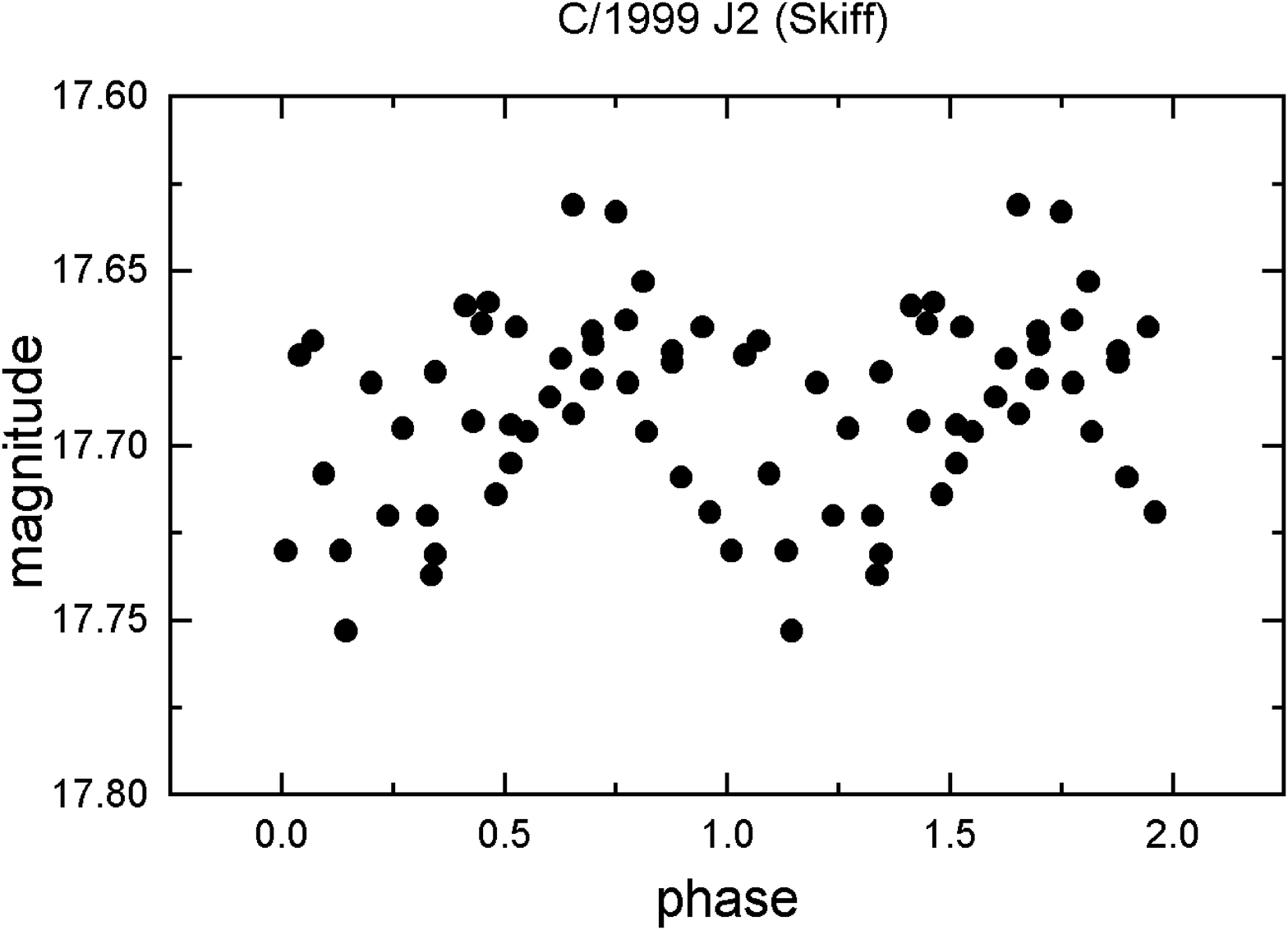,width=8cm}
\caption{The phase diagram of C/1999 J2 (Skiff) (P=1 h).}
\end{center}
\label{10298f4}
\end{figure}

\bigskip
\noindent
{\it C/1999 J2 (Skiff)}
\newline
\noindent
The object was detected for the first time on CCD frames of
the LONEOS-Schmidt (59 cm) telescope on May 13, 1999 for the first time
(Skiff et al. 1999).
This comet with a discovery brightness
of 16\fm0 has the third largest perihelion distance known
(7.110 AU, the transit was on April 5, 2000).
The apparent total visual brightness was 14\fm5
during the observations, and because of its high declination it was
a favourable target for observers in the northern hemisphere.
The derived absolute brightness is 3\fm0, which is quite
bright compared to other similar comets.
The intrinsic peculiarity of this comet was also
suggested by the dust antitail reported by Fukushima et al. (2000)
in May, 2000.

The observed V$-$R colour index of the inner coma is 0\fm25$\pm$0\fm05.
A weak ion tail is barely visible on the direct images, and therefore
the relatively bluish colour may be attributed to possible
C$_2$ emission in the V-band.
The obtained coma-corrected absolute photometry gave the following mean
nuclear brightness: $\langle R \rangle=19\fm9\pm1\fm0$.
The coma contamination was estimated to be 87$\pm6$\% of the total inner
intensity (formal error). Assuming 0.04 albedo, the calculated cross section
is $p_R C=4\pm3\ {\rm km}^2$ resulting in a nuclear diameter of $10\pm8$ km.
This is a quite large value, however, it is simply necessary to
support the tremendous activity observed. The logarithmic
coma brightness density is linear in the inner 13\farcs0 with a slope
of $-1.6\pm0.1$. This value is significantly larger than the expected
one for an isotropic steady-state coma and suggests strong interaction
of the outflow and the radiation pressure. 

Time-series observations showed that there were rapid small-amplitude
variations on a time-scale of an hour, though with small
significance. We present
the lightcurves in Fig.\ 3 (the data regard to 1\farcs3 aperture
radius). In order to quantify the cyclic variation, a standard
Fourier-analysis was performed. A very short period of $0\fh96\pm0\fh07$ 
was revealed, the phase diagram is plotted in Fig.\ 4.

As rotation is the easiest way to explain light variability,
we have compared the observed behaviour with the
rotational breakup calculations of Davidsson (1999).
Accepting rotational variability, the period of rotation is twice
the period of the lightcurves. This means approximately
2 hours for the rotational period, which is physically permitted
for thopse bodies which are smaller than 4 km in spherical
approximation.
This does not contradict the estimated diameter of the nucleus.
However, other alternative mechanisms cannot be exluded, which are
presently unknown. It is worthwhile noting that
a similarly fast oscillation has been found for the asteroid
1689 Floris-Jan (Pych 1999), which has long rotational period.
The rapid variations were suggested to be caused by secondary
rotational effects, though no firm physical explanation was drawn.

\begin{figure}
\begin{center}
\leavevmode
\psfig{figure=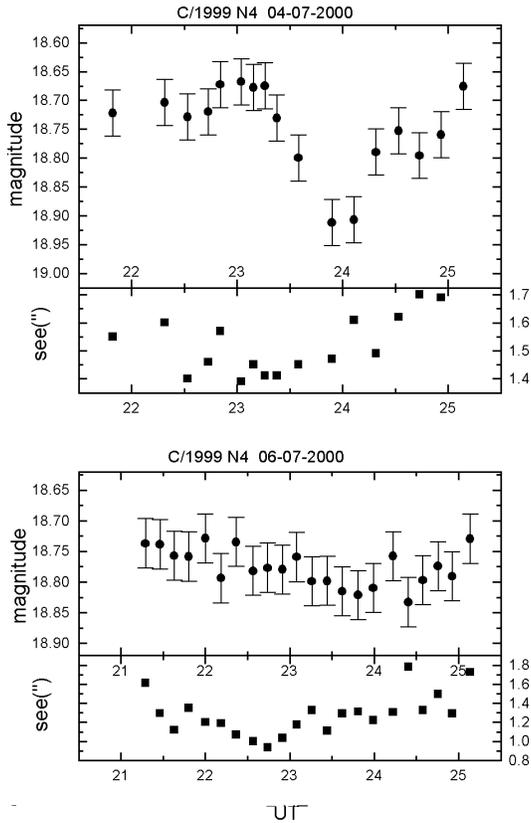,width=8cm}
\caption{The lightcurve of C/1999 N4 (LINEAR) on July 4 and July 6, 2000.}
\end{center}
\label{10298f5}
\end{figure}

\bigskip
\noindent
{\it C/1999 N4 (LINEAR)}
\newline
\noindent
This retrograde comet was found as an asteroid
by the LINEAR project on July 12, 1999 and its unusual
motion raised the question of its real nature
(Tichy et al. 1999). The discovery was made almost a year before
the perihelion at 5.505 AU, therefore, the evolution of this object
could be well monitored. The initial brightness of 17\fm5 
brightened up to 15\fm0.

Our measurements were taken on 2 nights. On July 4/5 the comet showed a light
variation of 0\fm3, the most striking feature of which is the rapid dimming
between 23.2 -- 01 UT. The lightcurve is presented in Fig.\ 5.
On the second night (July 6) we could detect only an ambiguous
variation with an amplitude not exceeding 0\fm08, while
a rotation effect would have been expected on this night too.
To exclude the correlation with the seeing, the seeing variation is
also presented below the lightcurves.

The V$-$R colour index of the inner coma is 0\fm47$\pm$0\fm05, fairly
close to the solar value ((V$-$R)$_\odot=0\fm36$, Meech et al. 1995).
This implies a relatively simple reflection with no emission and
little dust.
The obtained coma-corrected absolute photometry gave the following mean
nuclear brightness: $\langle R \rangle=20\fm6\pm0\fm2$.
The coma contamination was estimated to be 90$\pm5$\% of the total inner
intensity (formal error). Assuming 0.04 albedo, the calculated cross section
is $p_R C=0.4\pm0.3\ {\rm km}^2$ resulting in a nuclear diameter of $3\pm2$ km.
The logarithmic surface
brightness profile is the same as for the coma of C/1999 J2: a linear relation
in the inner 8\farcs0 with a slope of $-1.7\pm0.1$.

\begin{figure}
\begin{center}
\leavevmode
\psfig{figure=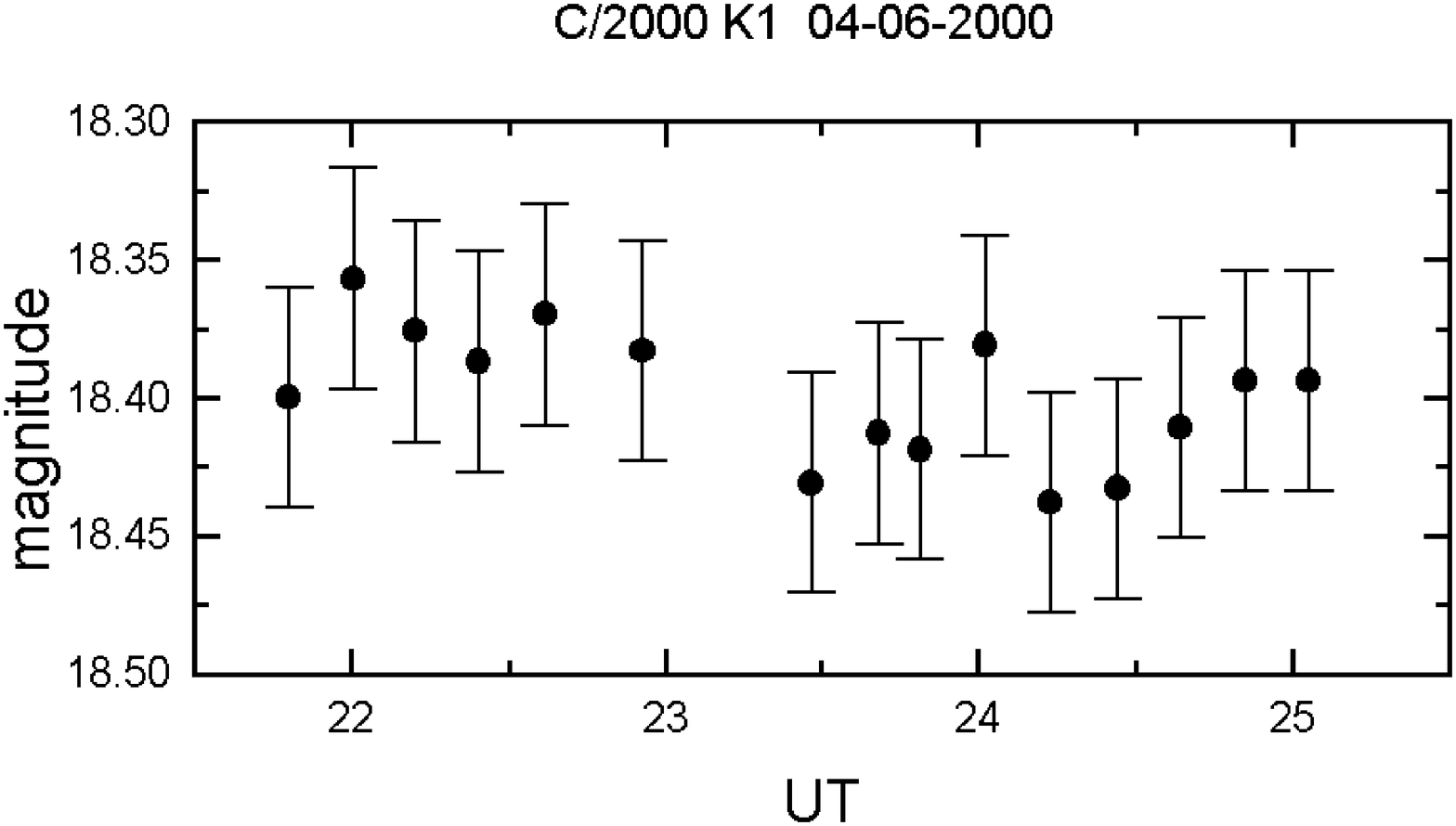,width=8cm}
\caption{The lightcurve of C/2000 K1 (LINEAR) on July 4, 2000.}
\end{center}
\label{10298f6}
\end{figure}

\bigskip
\noindent
{\it C/2000 K1}
\newline
\noindent
The retrograde comet was discovered at a brightness of 18\fm0
by the LINEAR project in June 2000 (Shelly et al. 2000).
Further prediscovery images
were found on frames from
the previous year (Amburgey \& Zoltowski 2000).
The perihelion passage occurred at 6.276 AU and
a visual brightness of 14\fm5.

We have found no significant variation during the observing run (Fig.\ 6).
Unfortunately, we could obtain only the presented 3-hour data series
which does not allow any firm conclusion to be drawn.
The V$-$R colour index of the inner coma is 0\fm68$\pm$0\fm05,
slightly reddish. This means a considerably dusty coma.
The obtained coma-corrected absolute photometry gave the following mean
nuclear brightness: $\langle R \rangle=19\fm5\pm1\fm0$.
The coma contamination was estimated to be 84$\pm10$\% of the total inner
intensity (formal error). Assuming 0.04 albedo, the
calculated cross section
is $p_R C=5\pm3 {\rm km}^2$ resulting in a nuclear diameter of
$11\pm8$ km. As for C/1999 J2, the remarkable tail and
coma activity requires a large nucleus.
The recorded logarithmic surface brightness profile has a slope
of $-1.55\pm0.1$ determined in the inner 14\farcs0.

\section{Discussion}

The presented observations revealed some basic parameters
of three interesting distant comets. Despite the large heliocentric
distances of 5.5--7.2 AU, we could observe extended tails
and fairly bright comae. The canonical view of
cometary activity at such distances excludes the
possibility of water ice sublimation as the inner
``engine'' of the gas and dust production, because the
heat from the Sun is not sufficient. Usually, it is assumed
that a CO- or CO$_2$-dominated nucleus may be the
source of this activity (Houpis \& Mendis 1981, Luu 1993).
Bar-Nun \& Prialnik (1988) discussed the possible
hydrogen comae in distant comets as a further
mechanism beyond $\sim$4 AU.
Based on a very extensive
database of comet observations
($\sim$50 comets over a range of 1 to $\sim$30 AU),
Meech \& Hainaut (1997) have clearly shown that
dynamically young comets are intrinsically brighter,
exhibiting dust comae and activity at large distances
in the region where water ice sublimation is
not possible. Our observations fit very well the
description above.

Because of the unexpectedly strong coma contamination, the presented
results may suffer from large systematic errors, as
the applied coma subtraction is likely an
oversimplification of the real case. Thus, the
calculated nuclear diameters should be considered as
well-educated guesses with possibly large systematic
uncertainties (up to 5--10 km).
The recorded surface brightness
profiles with logarithmic derivatives $-$1.6..$-$1.8
are characteristic for isotropic steady-state outflows
affected by the radiation pressure alone (Jewitt 1991, Luu 1993).

Finally, the main observational results can be summarized as follows:

\noindent 1. CCD observation of three distant comets taken
on three night in July, 2000 are presented. The time-series
data revealed short-term variations in two comets, while the
third one was constant at a level of $\pm$0\fm05. As the
time span is quite short, no firm conclusion was drawn
on the possible rotation.

\noindent 2. The absolute photometry was corrected
for the presence of comae and nuclear radii were
determined assuming a 0.04 albedo. The results are:
C/1999 J2 -- 10 km, C/1999 N4 -- 3 km, C/2000 K1 -- 11 km.

\noindent 3. The observed V$-$R colours imply a slightly
reddish coma for C/2000 K1, a normal solar reflected light for
C/1999 N4 and a slightly bluish coma for C/1999 J2. The latter one
may be caused by the C$_2$ emission decreasing the
V-magnitude. The ion tail of C/1999 J2 was also detected.

\noindent 4. We note that our results are strongly affected by
selection effects, as the target list was formed to include
the brightest and most distant active comets
visible in July, 2000.

\noindent 5. As a by-product of this project, coma and tail
characteristics of three other comets have
been also examined. The results are also discussed, although two of these
targets were not further than 4 AU.

\bigskip
\noindent

\begin{acknowledgements}
This research was supported by
Pro Renovanda Cultura Hungariae Grants DT 2000. m\'aj./43.,
DT 2000. m\'aj./44., DT 2000. m\'aj./48.
and DT 1999. \'apr./23, FKFP Grant 0010/2001, OTKA Grant \#T032258,
``Bolyai J\'anos'' Research
Scholarship of LLK from the Hungarian Academy of Sciences
and Szeged Observatory Foundation.
The referee, Dr. Olivier Hainaut, helped to improve
the paper with his comments and suggestions.
The warm hospitality and helpful assistance 
of the staff of Calar Alto Observatory is gratefully acknowledged.
The NASA ADS Abstract Service was used to access data and references.
\end{acknowledgements}

\end{document}